\documentclass[aps,preprint]{revtex4}

\usepackage{graphics,epsfig,amssymb}

\usepackage{amsmath}

\def\be{\begin{equation}}
\def\ee{\end{equation}}

\begin{document}
\title{ Excited states of zero seniority  based on a  pair condensate}

\author{Th. Popa and  N. Sandulescu\footnote{coresponding author (email:sandulescu@theory.nipne.ro)}}
\affiliation{
National Institute of Physics and Nuclear Engineering, 76900 Bucharest, Romania}
\author{M. Sambataro}
\affiliation{ Istituto Nazionale di Fisica Nucleare - Sezione di Catania, Via S. Sofia 64, I-95123 Catania, Italy}
 
\begin{abstract}

We study the excited states of zero seniority for various like-particle systems 
interacting by pairing forces and by general two-body interactions.
We consider two types of excitations, generated from a ground state described by a pair
condensate. One type is obtained by breaking a pair from the ground state condensate and
replacing it by "excited" collective pairs built on time-reversed single-particle orbits. 
The second type of zero seniority excited states is described
by a condensate of identical excited pairs. The structure of these excited states is
analysed for the picked fence model and for the valence neutrons of $^{108}$Sn. For a state-depending
pairing interaction, the first type of excited states agree well with the J=0 states which are
known in $^{108}$Sn. At the same time, these states can be also associated
unambiguously with those exact states which are the closest in energy to the experimental
levels. The states corresponding to the excited pair condensate appear at low energies, around the energy
of the second excited state of the first type, and they do not have
a simple correspondence with exact eigenstates. However, at a much higher excitation energy there is an
exact state which is similar in structure to an EPC state. It is shown that this EPC state has
the features of a giant pairing vibration.  

\end{abstract}

\maketitle

\section{Introduction}

As pointed out  many years ago \cite{bohr}, basic nuclear properties can be simply explained in 
the framework of BCS approach \cite{bcs}. Presently, BCS-like approximations are employed
to treat the pairing correlations in almost all mean-field calculations based on realistic 
energy density functionals (EDF). The main drawback of these calculations is the
fact that they break the particle number conservation. Restoring the right number of 
particle in the EDF calculations  is not a trivial task \cite{robledo}. 
The simplest alternative, which can be applied to any Hamiltonian in which the mean
field and the pairing are decoupled, is to use from the beginning not the BCS but 
the particle number projected-BCS (PBCS) approach \cite{bayman,blatt,talmi}. We remind that in the PBCS
approach the ground state is approximated by a pair condensate in which the number of pairs
is fixed by the actual number of particles involved in the pairing calculations. 
The accuracy of BCS and PBCS approximations in describing ground state properties of
like-particle systems interacting by pairing forces has been analysed in several studies. 
The main conclusions are the following: 
(a) BCS underestimates significantly the pairing correlations energies for 
finite systems,  even in the limit of strong coupling \cite{richardson}; 
(b) PBCS provides much better results \cite{dietrich}, especially when it is applied 
in model spaces of the order of one major shell \cite{sandulescu_bertsch}.
    
The BCS approach provides a simple scheme for understanding not only the ground state 
properties of nuclei but also the pattern of their excitation spectra. In BCS 
the excited states are associated to quasiparticles generated  by breaking particles
from the BCS condensate and distributing them in single-particles states which 
are not participating to the pairing correlations.  This is a rather crude approximation because it does not take 
into account the interaction between the quasiparticles and also because these excited states
do not have a well-defined particle number. In addition, as pointed out already
many years ago \cite{richardson}, in the BCS approximation one cannot unambiguously define
the excited states of zero seniority. A better alternative is to start with the PBCS ansatz
for the ground state and to construct excited states by breaking pairs from the PBCS condensate and replacing  them by "excited" collective pairs (e.g., see \cite{talmi,gambir,bonsignori,allaart} 
and the references quoted therein). In this study we shall focus on a particular type of these
PBCS-based excited states, namely the states of zero seniority. Apart from one-broken-pair states
we shall also study another type of excited states of zero 
seniority, which, according to our knowledge, has not been considered before. These excited
states have the structure of a pair condensate built by identical excited pairs. The two types of zero
seniority excited states will be analysed for various systems and their accuracy will be probed 
with Hamiltonians which can be solved exactly. 

\section{Formalism} 

In the present study we consider systems formed by an even number of spin-$1/2$ fermions, 
e.g., neutrons or protons, distributed on $\Omega$ single-particle states $|i\rangle$ of 
energies $\epsilon_i$ and interacting by a two-body force. For a force of pairing type
these systems are described by the Hamiltonian
\begin{equation}
\label{hbcs}
H= \sum_i^\Omega \epsilon_i (a^\dagger_i a_i +a_{\bar i}^\dagger a_{\bar i}) +
  \frac{1}{4}\sum_{i,j=1}^\Omega v_{ij} a^\dagger_i a^\dagger_{\bar{i}} a_{\bar{j}}a_j. 
\end{equation}
In the equation above $\bar{i}$ denotes the time-reversed state $|\bar{i} \rangle = \hat{T} |i \rangle$ 
and $v_{ij} \equiv \langle i\bar{i}|\hat{V}|j\bar{j} - \bar{j}j \rangle $.

Usually, the ground and excited states of Hamiltonian (1) are  approximated by employing
BCS-based models. Here we shall focus on zero seniority excited states
which can be generated starting from the PBCS condensate. These states, introduced in the
subsection (A), will be compared with the exact solutions of the Hamiltonian (1). For the
state-independent pairing forces, i.e., $g=v_{ij}$, we shall use  the exact analytical
solutions found by Richardson many years ago \cite{richardson}. The Richardson solutions, and
their relations to the PBCS-based states,  are presented shortly in the subsection (B).

\subsection{Zero seniority states based on the PBCS condensate}

In the BCS approximation the ground state of Hamiltonian (1) is expressed
as  a coherent superposition of Cooper pairs, i.e.,
\begin{equation}
| BCS > \propto  e^{\Gamma^\dagger} | - > \equiv  \sum_n
\frac{(\Gamma^\dagger)^n}{n!} | - > , 
\end{equation}
where the Cooper pairs are defined by
\begin{equation}
\Gamma^\dagger = \sum_i x_i a^\dagger_i a^\dagger_{\bar{i}} .
\end{equation}
The mixing amplitudes of the pair operator are usually written as $x_i=v_i/u_i$, where the parameters
$v_i$ and $u_i$ are determined by solving the well-known BCS equations.
 
The particle number projected-PBCS (PBCS) approximation is obtained by restricting the 
expansion in Eq.(2) to the term having the required number of pairs $N$. 
Thus, in PBCS  the ansatz for the ground state is the pair condensate
\begin{equation}
| PBCS > = (\Gamma^\dagger)^N | - > .
\end{equation}
In the majority of applications the pairing is supposed to act only 
in a finite region around the Fermi level. In this case $N$ denotes only the particles
involved in the pairing correlations,  while the other particles of the system are included 
in the "vacuum" state  $|- \rangle$.

The mixing amplitudes $x_i$ of the pair operator (3) are determind variationally by minimizing
the average of the Hamiltonian in the PBCS state (4) and imposing the normalisation condition 
$\langle PBCS|PBCS \rangle =1$. The corresponding PBCS equations satisfied by the 
parameters $x_i$ can be found by employing the residual integrals technique \cite{dietrich}. 
Alternatively, if the number of pairs is not too large, the amplitudes $x_i$ can be obtained by 
using the recurrence relations method \cite{pbcs_t1}. These relations are given in the Appendix.

In the framework of PBCS approach the excited states are usually obtained by breaking pairs from the 
PBCS condensate and contructing new collective "excited" pairs which are attached to the remaining 
PBCS condensate \cite{talmi,gambir,bonsignori,allaart}. In the present study we are interested in
those excited states in which the excited collective pairs are built from pairs of fermions 
distributed in time reversed orbits. For the case 
of one-broken-pair approximation these excited states have the expression 
\begin{equation}
|N;1_k > = \tilde{\Gamma}_k^\dagger (\bar{\Gamma}^\dagger)^{N-1}| 0 >
\end{equation}
where  
\begin{equation}
\bar{\Gamma}^\dagger = \sum_i y_i a^\dagger_i a^\dagger_{\bar{i}} 
\end{equation}
and
\begin{equation}
\tilde{\Gamma}_k^\dagger = \sum_i z^{(k)}_i a^\dagger_i a^\dagger_{\bar{i}} 
\end{equation}

The mixing amplitudes $y_i$ and $z^{(k)}_i$ are determined variationally by minimizing the average of the
Hamiltonian on the state (5) under the constraints $ \langle N;1_k| N;1_{k'} \rangle =\delta_{k,k'}$ and
$\langle N:1_k|PBCS \rangle = 0$.
It can be shown that these conditions are satisfied if the amplitudes
$y^{(k)}_i$ and $z^{(k)}_i$ are satisfying the recurrence relations given in the Appendix . 

In the same manner one can construct excited states with more broken pairs. Of special interest for this study
are the excited states in which all the pairs are broken and replaced by a unique excited collective pair.
These excited pair condensate (EPC) states have the expression
\begin{equation}
|EPC(k) \rangle  = (\hat{\Gamma}^\dagger_k)^N | 0 >
\end{equation}
where $\hat{\Gamma}_k^\dagger = \sum_i w^{(k)}_i a^\dagger_i a^\dagger_{\bar{i}}$. The mixing amplitudes
$w^{(k)}_i$ are determined variationally from the minimisation of the average of the Hamiltonian on 
the state (8) under the contraints $ \langle EPC(k)| EPC(k') \rangle =\delta_{k,k'}$ and 
$\langle EPC(k)|PBCS \rangle = 0$.  The calculation scheme for 
the EPC states is similar to the one employed for the ground PBCS state. 

As it can be noticed, in all the states introduced above the collective pairs are built as a superposition of 
two particles configurations  $a^+_i a^+_{\bar{i}}$. Due to this reason, these states are
referred to as zero seniority states.  If the single-particle states are spherically symmetric, 
both the ground state and the excited states of the system defined by Eqs. (4,5,8) have the
angular momentum $J=0$. 

\subsection{Zero seniority states and the Richardson solution of state-independent pairing}

As shown in Ref. \cite{richardson}, for a state-independent pairing interaction,
the pairing Hamiltonian (1) can be solved analytically.  For an even number of particles,
the exact solution for the ground state has the expression \cite{richardson}
\begin{equation}
| \Psi > = \prod_\nu^N B^\dagger_\nu | 0 >,
\end{equation}
where the pair operators  $B^\dagger_\nu$ are defined by
\begin{equation}
B^\dagger_{\nu} = \sum_i \frac{1}{ 2\varepsilon_i-E_{\nu} } a^\dagger_i
a^\dagger_{\bar{i}} . 
\end{equation}
They depend on the parameters $E_\nu$ which satisfy
the set of non-linear equations
\begin{equation}
\frac{1}{g} - \sum_j \frac{1}{2\varepsilon_j-E_\nu}
+\sum_{\mu \not= \nu} \frac{2}{E_\mu-E_\nu}
 = 0 .
\end{equation}
The sum of the parameters $E_\nu$ gives the total energy of the system, i.e.,
\begin{equation}
E = \sum_\nu E_\nu .
\end{equation}
In the limit $g=0$ the pair energies $E_\nu$ associated with the ground state
coincide with twice the lowest single-particle energies, i.e.,
$E_\nu=2\varepsilon_\nu, (\nu=1,2, ...N)$, where $N$ is the number of pairs.
At specific finite values of the interaction strength,  the pair energies turn two by
two from real to complex conjugate quantities.

There are two kinds of excited states which can be contructed starting from the ground
state (9). One kind is obtained by breaking pairs and distributing the corresponding
particles on different single-particle states, which will not participate in the pairing
correlations. These excitations are the analogous of BCS states of non-zero seniority.
The second kind of excited states, corresponding to zero seniority states, are obtained
by keeping the number of collective pairs unchanged and modifying the initial conditions for the
the energies $E_\nu$ in the limit $g=0$. For example,
the lowest excited state corresponds, in the limit $g=0$, to $E_\nu=2\epsilon_\nu$
for $\nu=1,2,..N-1$ and $E_N = 2\epsilon_{N+1}$.
In Ref.\cite{richardson} this excited state is labeled as $(-1)^2(+1)^2$, indicating that this state is
obtained, in the linit $g=0$, by promoting a pair from the last occupied level to the
next unoccupied level. By this procedure one can  generate excited 
states which have, formally, the same structure as the ground state (9), but expressed
in terms of new "excited" $B_\nu$ pairs. The latter are defined by new parameters $E_\nu$ which 
are determined by solving the Richardson equations (11) with the new initial conditions 
at $g=0$.

As it can be seen, the exact and the PBCS-based states are quite different from each other.
Contrasting them allows to evidence the limits of the PBCS  approximations. 
For the PBCS-based states it is assumed that all the pairs $B^+_\nu$ of the exact solution, except of an
"excited" pair in the case of the states (5),  can be represented by a unique collective 
pair $\Gamma^+$, which is supposed to average out the effect of the distinct pairs $B^+_\nu$. 
This is a reasonable approximation for those pairs $B^+_\nu$ which have similar structures.
These are the pairs  with the parameters $E_\nu$ not too far from the Fermi level and not
too close to a single-particle energy. Consequently, the PBCS-based  approximations are
expected to work  well if they are applied in a finite region around the Fermi level and 
for a rather strong pairing strength, typically greater than the critical BCS value. 

Another difference between the exact and the PBCS-based solutions is in the number of parameters
which defines them. In the case of the exact solution the number of
parameters, $E_\nu$, is equal to the number of pairs. On the other hand, for the PBCS-based 
states (4,8) the number of parameters is equal to the number of single-particle orbits included
in the model space and twice this number for the states (5) . Therefore, the accuracy
of the PBCS-based approximations  is expected to be better for large model spaces. 

A specific feature of the exact solution, which is absent in the PBCS-based 
approximations, is the correlations between the $B^+_\nu$ pairs. These correlations
appear when some of the parameters $E_\nu$ become complex, which happens beyond
a critical value of the pairing strength. In this case, the two pairs $B^+_\nu$ 
corresponding to the complex conjugate parameters ($E_\nu,E^*_\nu$) are entangled. 
More precisely, they form correlated 4-body structures \cite{sasa}. How important
are these 4-body structures in nuclei is still an open issue.

\section{Accuracy of excited states of PBCS-type} 

To probe how well the approximations (5) and (8) work for the excited states,
we shall consider first the case of a state-independent pairing Hamiltonian, 
which can be solved exactly by using the Richardson method. Then we shall 
analyse the PBCS-based states for a state-dependent pairing interaction and
a general two-body force of shell-model type.

\subsection{Excited states of zero seniority for a state-independent pairing force}

\begin{figure}[h]
\centering
\includegraphics[width=0.50\textwidth, angle=-90]{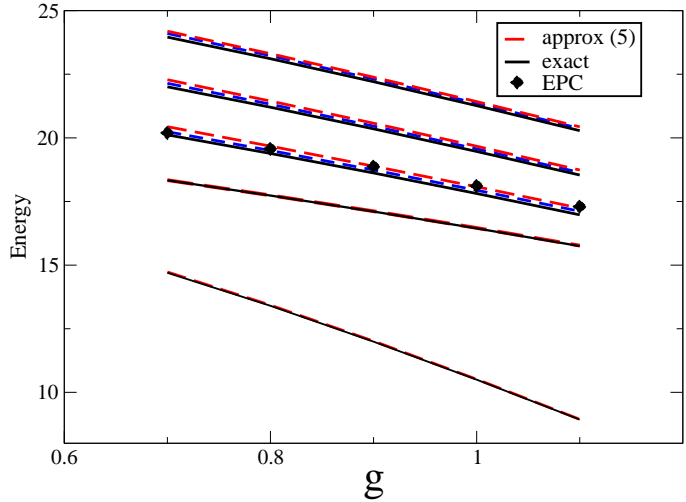}
\caption{The energies of the states (4,5) compared to the exact results 
as a function of pairing strength. By full triangles we show the energies 
of the excited pair condensate (8). The results correspond to 4 pairs distributed
in 8 equidistant and doubly degenerate states.}
\end{figure}

\begin{figure}[h!]
\centering
\includegraphics[width=0.70\textwidth, angle=-90]{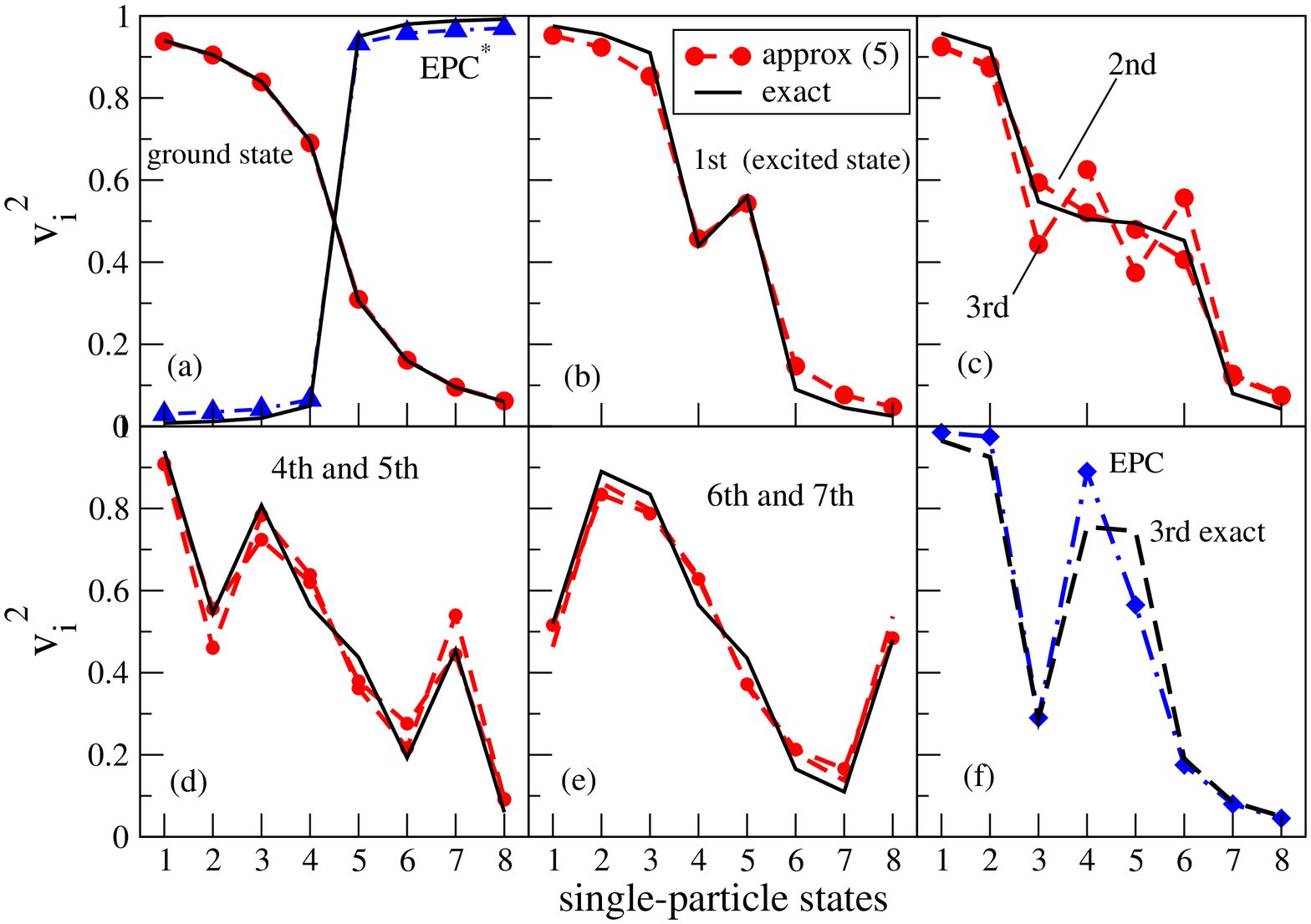}
\caption{ Occupation probabilities of single-particle states for the states (4,5,8)
and the exact states which are closest in energy. For the exact degenerate 
states we show the sum of the occupation probabilities. For the states EPC$^*$ 
and EPC shown in panels (a,f) see the text. The results are for the pairing strength g=0.7.}
\end{figure}

We  start by investigating the excited states for the systems 
formed by an even number of particles distributed in a set of equidistant
and doubly degenerate levels. More precisely, we consider $\Omega$ levels,
of energies $\epsilon_i$ =1,2, ...$\Omega$,
and a number of particles corresponding to half-filling, i.e., $2N=\Omega$,
where $N$ is the number of pairs. These systems, with $N=4-16$, have been
considered by Richardson in order to analyse the performance of the BCS approximation
relative to the exact solution \cite{richardson}. Similar systems, with $N=4-40$, 
have been used later on to analyse the accuracy of the PBCS approximation 
for the ground state \cite{sandulescu_bertsch}.

Here we shall take as an example a system with $N=4$ and $\Omega=8$. As for the pairing strength,
we use the same range employed by Richardson, i.e., $g=0.7-1.1$. For this system there are 
8 zero-seniority states of the form (5). The energies of these states are shown in Fig. 1. 
For the lowest state the collective pairs (6) and (7) have a similar structure and both of them
are similar to the ground state pair $\Gamma^+$ of the PBCS condensate (4). Therefore the state (5) 
with the lowest energy shown in Fig. 1 corresponds  to the PBCS ground state. 

The first excited state (5) is calculated by variationally determining both the pair $\bar{\Gamma}^+$ of 
the broken condensate and the excited pair. A simpler approximation, which we apply for the higher 
excited states (5), is to keep for  $\bar{\Gamma}^+$ the same structure as for the ground state.

For comparison, Fig. 1  also shows the exact energies for the lowest 8 states of zero seniority. 
The first energy corresponds to the ground state while the second to the Richardson solution 
denoted by $(-1)^{2}(+1)^2$, i.e., a state which is obtained, in the limit $g$=0, by promoting a
pair from the last occupied level to the first unoccupied level. The next energies correspond
to doubly degenerate solutions. For example, the 3rd energy corresponds to the 3rd and the 4th
degenerate states. In the limit $g=0$ these states correspond to the configurations
$(-1)^{2}(+2)^{2}$ and $(-2)^{2}(+1)^{2}$, which have the same energies.  

From Fig. 1 one can notice that the ground state energy predicted by PBCS is very accurate.  
Indeed, the maximum deviation from the exact value, corresponding to $g$=1.1, is 
about 0.26$\%$. 
The energies of excited states (5) follow rather closely the exact energies. In particular one can see 
that to the exact doubly degenerate states correspond to two non-degenerate states (5) with energies close
to the exact value. 

The energy of the lowest excited pair condensate (EPC) is shown in Fig.1 by full triangles.
It can be seen that the EPC energy follows closely the exact energy  of the 3rd and 
the 4th degenerate states. The energy of the EPC state is unexpectedly low, taking into account
the fact that this state is obtained by breaking all the pairs from the ground state
condensate. 

An indication of how close the wave functions (4,5,8) are to the exact states is obtained from
the comparison between the predicted occupation probabilities of the single-particle orbits.
This comparison is done here for the strength g=0.7. The occupation probabilities $v^2_i$ 
corresponding the the ground state (4) and the first excited state (5) are shown in Fig. 2(a,b). 
As it can be seen, the $v^2_i$ for these states follow the exact results rather closely, 
especially for the ground state (4).

A comparison with the exact wave functions corresponding to the doubly degenerate states is not straightforward. 
This is because any combination of the two degenerate solutions is
also exact eigenfunction of the Hamiltonian. 
Due to this reason the occupation probabilities $v^2_{i}(1)$  and $v^{2}_{i}(2)$ associated with the two degenerate
states are not well-defined because they depend on the chosen representation for these states. What does not
depend on the representation is the sum 
$v^2_{i}(12) = v^2_{i}(1)+v^2_{i}(2)$. In Fig. 2(c,d,e) one can see the comparison between $v^2_{i}(12)$ 
and $v^2_i$ associated to the two states (5) which are closest in energy to
the exact degenerate states. It can be seen that $v^2_i$
and $v^2_{i}(12)$ are rather similar.

The occupation probabilities $v^2_i$ corresponding to the EPC state are shown in Fig. 2(f).
In the limit $g=0$ the EPC state corresponds to the configuration $(-1)^{2}(+2)^{2}$. Since, in the limit $g=0$, this  is also the configuration 
of the 3rd exact state, we expect that the latter to be closer to the EPC state. In order to check that,
we have done another calculation in which we have removed the degeneracy of the exact states by adding a
very small perturbation. The $v^2_i$ corresponding to the 3rd new state are shown in Fig. 2(f) 
by dashed-dotted line. One can notice that the $v^2_i$ corresponding to  the exact state and to  the 
EPC state are rather different for the 4th and the 5th orbits.

An interesting question is whether there is an exact eigenstate of the Hamiltonian which is
similar in structure to a EPC state. To answer this question, for each exact eigenstate 
$|\Psi \rangle$ of zero seniority we determined the EPC state which maximises the overlap 
$\langle EPC | \Psi \rangle $ and which is orthogonal to the ground state pair condensate.
We found that the eigenstate $|\Psi \rangle$ with the highest energy, equal to -49.935, 
is the one that resembles the most a EPC state. The overlap of this eigenstate with 
the corresponding EPC state, called below the EPC$^*$ state, is equal to 0.986. 
The energy of the EPC$^*$ state
is equal to -49.596, a value rather close to the exact energy of the $|\Psi \rangle$ state. 
The occupation probabilities $v^2_i$ of the single-particle orbits 
for the EPC$^*$ state are shown in Fig. 2(a). As can be seen, they follow closely the exact values.
By contrast to the EPC state shown in Fig. 2(f), the occupation probabilities for the  EPC$^*$ 
state have a smooth dependence on single-particle energies, as in the case of the ground state condensate. 
As seen from Fig. 2(a), the EPC$^*$ state looks as the reverse of the ground 
state condensate, in the sense that in the two states the role of the lower
and higher energy orbitals is interchanged.

\subsection{Excited states of zero seniority for a state-dependent pairing force}

To benchmark the accuracy of the  states (4,5,8)  in the case of  state-dependent pairing forces, 
we consider a pairing interaction derived from G-matrix calculations \cite{g-matrix}. Its matrix elements and the 
energies of the single-particle states are given in Ref. \cite{volya}.
With this interaction we have calculated  the seniority zero states for $^{108}$Sn,
taking $^{100}$Sn as core. 

\begin{figure}[h]
\centering
\includegraphics[width=0.55\textwidth, angle=-90]{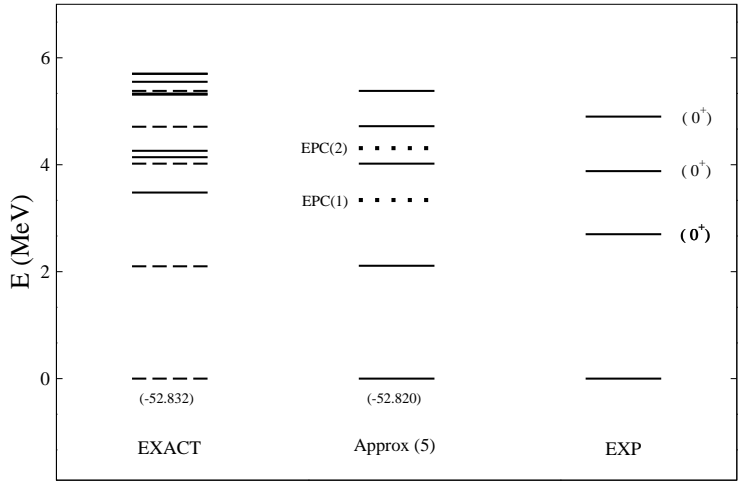}
\caption{ Energies corresponding to the states (4,5,8) compared to the exact spectrum 
and experimental energies \cite{exp}. The exact energies which correspond
to the  states (4,5) are indicated by dashed line. EPC values are the energies of the excited pair condensate. All calculated levels 
have $J=0$. The results are for the neutrons in the valence shell of $^{108}$Sn interacting through
a state-dependent pairing interaction.  }
\end{figure}
\begin{figure}[h]
\centering
\includegraphics[width=0.70\textwidth, angle=-90]{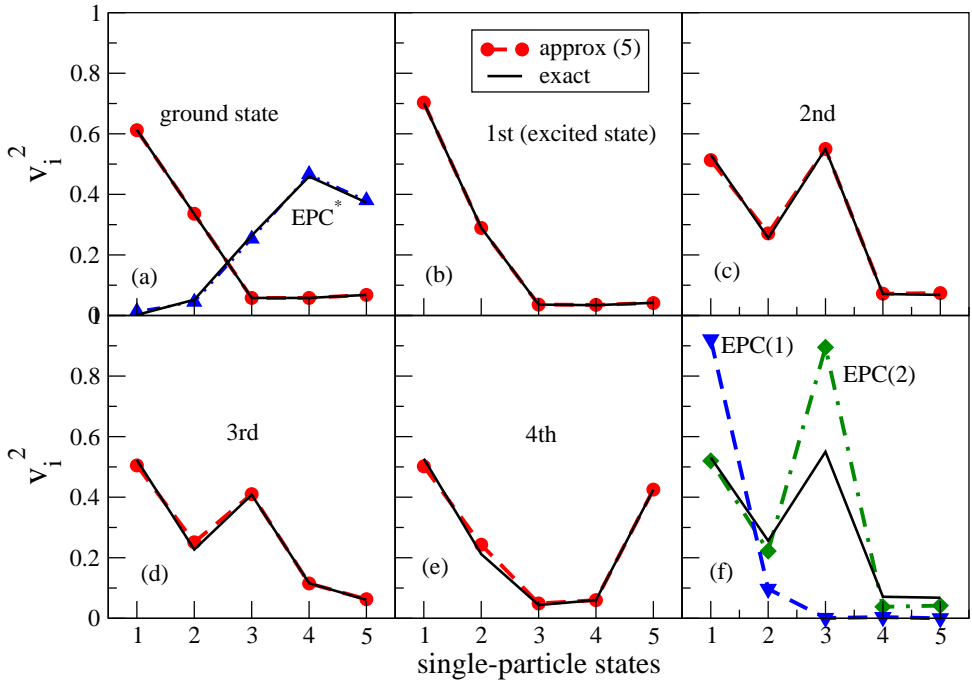}
\caption{ Occupation probabilities of single-particle (s.p.) orbits corresponding to 
the states (4,5,8) compared to the exact states which have the 
closest similarity with the former. For the states EPC$^*$ and EPC(1,2) shown in panels
(a,f), see the text.  The five s.p. states on the x-axis 
are, from the left to the right, $1g_{7/2}$, $2d_{5/2}$, $3s_{1/2}$, 
$1h_{11/2}$ and $2d_{3/2}$. The results are for the neutrons in the 
valence shell of $^{108}$Sn interacting through  a state-depending pairing interaction.}
\end{figure}
As in the previous section, the first excited state (5) is calculated  by variationally
determining both  the excited pair and the pair $\bar{\Gamma}^+$ which defined the
broken pair condensate. For the other excited states we determine variationally only the
excited pair while for the pair $\bar{\Gamma}^+$ we take the same structure as for
the ground state condensate. 

The predictions of the states (4,5,8) for the energies are shown in Fig. 3. 
In the same figure we show the energies of the exact states with $J=0$ obtained by
diagonalization. The exact energies which correspond
to the states (4,5) are indicated by dashed lines. This correspondence is supported by the very good agreement between
the occupation probabilities of the orbits, shown in Fig. 4.
 
Fig. 3 also includes the  experimentally known levels which, most probably, have
the spin $J=0$. One can notice that the energies of these levels are well described by the
states (5). In this energy region there are at least 6 exact J=0 states. The exact 
states which agree better  with the experimental levels are the ones which correspond to 
the three lowest states (5). Consequently, it appears that the lowest three 
known $J=0$ states in $^{108}$Sn have a simple physical interpretation: they
have a structure of one-broken-pair type.

In Fig. 3 are shown  the energies of two EPC states (8). They correspond to the first two
minima obtained variationally with the state (8). The occupation probabilities of the single-particle
states corresponding to the two EPC states are shown in Fig 4(f). The lowest EPC state is practically
built on the first two single-particle orbitals, $g_{7/2}$ and $d_{5/2}$. On the other hand,
the second EPC state is spread out on all the orbits. In the limit of small pairing strength
the two EPC states correspond to three pairs promoted from the orbit $f_{7/2}$ to the 
orbit $d_{5/2}$ and, respectively, to one pair promoted from $f_{7/2}$ to $3s_{1/2}$. 
In the Richardson notations these configurations are $(-1)^6 (+1)^6$ and $(-1)^2 (+3)^2$. 
In the energy region of the EPC states there are few exact states. In Fig. 4(f) we show
the comparison with the 3rd exact excited state, which has the closest similarity to
the EPC(2) state. It can be noticed that this exact state has a significantly lower
occupation probability for the orbit $3s_{1/2}$. 

As in the case of the picket fence model, we have searched whether there is an exact eigenstate
of the state-depending pairing Hamiltonian which is similar in structure to a EPC state. 
To find this state we have used the procedure we explained at the end of the previous section.
The search was done for the 49 exact seniority zero states obtained by diagonalising the
Hamiltonian in the space of pairs. We have found that the 46th state, of 
energy -31.854, resembles the most an EPC state. This eigenstate has the largest overlap
with an EPC state, equal to 0.995. The EPC state, which we shall denote by EPC$^*$, 
as in the previous section, has an energy equal to -31.905, which is very close to the
energy of the corresponding exact state. The occupation probabilities of the single-particle
orbits associated to the EPC$^*$ state are shown in Fig. 4(a). As can be seen, they can be hardly 
distinguished from the occupation probabilities of the corresponding exact state. From Fig. 4(a) one
can notice that the EPC$^*$ state is mainly built on the high energy orbits $1h_{11/2}$ and $2d_{3/2}$. 
This is the reversed situation compared to the ground state condensate, in which the low energy orbits
$1g_{7/2}$ and $2d_{5/2}$ have the highest occupation probabilities. 

In the context of tin isotopes, the EPC$^*$ state discussed above has  common features with
the so-called giant pairing vibration (GPV). For a recent overview of the theoretical and experimental 
work done on GPV see Ref. \cite{gpv}.  Here we recall that the GPV is usually defined 
as a collective excited state composed by a coherent superposition of 
particle-particle configurations, analogous to the ground state. However, contrary to the 
ground state, the GPV state is expected to be formed by particle-particle configuration 
built on the next major shell rather than the valence shell. This is the standard 
scenario for the formation of the GPV state, which was initially considered  for the case
of lead isotopes \cite{broglia,liotta}. The light tin isotopes, in particular 
the isotope $^{108}$Sn considered in this paper, offer an interesting alternative 
to the standard scenario mentioned above. As it is known, in tin isotopes the neutron 
number N=64 is a quasi-magic number. This is due to the fact
that the lowest two orbits $1g_{7/2}$ and $2d_{5/2}$ are rather well separated in energy
from the last two  orbits 
$1h_{11/2}$ and $2d_{3/2}$. In between these orbits there is the state $3s_{1/2}$, but, since
its degeneracy is small, this state does not affect much the quasi-magic character of N=64. 
Consequently, since in $^{108}$Sn the two first orbits play the role of the valence shell while
the last two orbits act as the next major shell, the  EPC$^*$ state discussed above has
the characteristic of a GPV state  built on the 
orbits $1h_{11/2}$ and $2d_{3/2}$.

\subsection{Excited states of zero seniority for a general shell-model-type interaction}

In order to analyse the seniority zero states for the case of  a general 
two-body force we consider as an example a shell-model interaction \cite{interaction} 
which was fitted in order to get a reasonable description of the low-lying states
of tin isotopes \cite{chong}. The one-broken-pair approximation was applied
for tin isotopes and realistic interactions in various studies 
\cite{gambir,bonsignori,allaart,sandulescu_gsen}, but in none of them 
the accuracy of this approximation was analysed for the excited states of 
zero seniority. Here we examine the states of zero seniority for the tin
isotope $^{108}$Sn and the predictions of the approximations (4,5,8) will
be compared to the exact solutions provided by exact diagonalisation. 
\begin{figure}[h]
\centering
\includegraphics[width=0.55\textwidth, angle=-90]{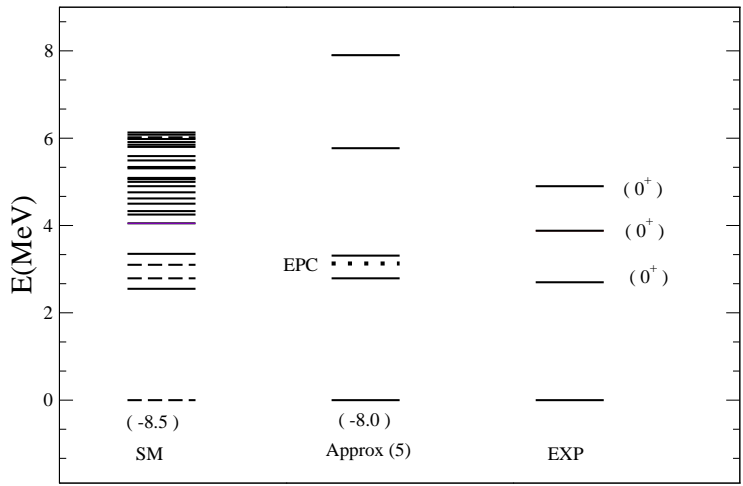}
\caption{The same as in Fig 3 but for a general two-body force of shell model type. }
\end{figure}
\begin{figure}[h]
\centering
\includegraphics[width=0.70\textwidth, angle=-90]{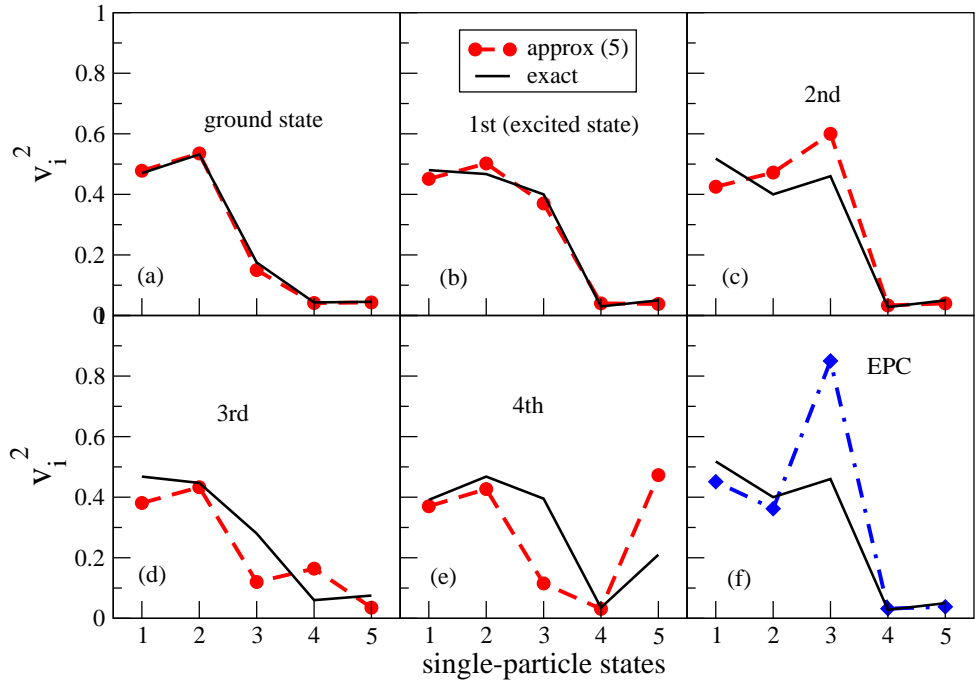}
\caption{The same as in Fig. 4 but for a general two-body force of shell model type.}
\end{figure}

All excited states (5) are calculated by replacing the pair $\bar{\Gamma}^+$, 
which defines the broken condensate, by the pair of the ground state condensate (4).
The energies of the states (4,5,8) are shown in Fig 5. In Fig. 5 we also show the
exact shell-model (SM)  energies, calculated with the code bigstick \cite{bigstick}. 
As expected, for the case of the general interaction the 
correspondence between the states (5) and the exact states is not so straightforward 
as for the pairing interactions. First of all, from the occupation probabilities shown in
Fig. 6 we observe that the SM ground state can be well approximated by a pair condensate.
There is also a very good agreement between the first excited state (5) and the second SM
excited state. The 2nd state (5) corresponds to the 4th SM excited state, but the agreement
between the occupation probabilities is not as good as for the first excited state.
In the energy region of the 3rd and the 4th excited states (5) there are many SM states,
which make the comparison difficult. In Fig. 6 we have shown the comparison with the 
SM states which we found to have the closest resemblance in occupation probabilities.
As it can be seen, for these states the differences are significant. 

From Fig. 5 one can notice that the PBCS approximation provides a binding energy which
is very close to the exact result. There is also a very good agreement between the 
energies of the second excited SM, the first one-broken-pair state and the first 
experimental J=0 level. Surprisingly, the agreement between the other experimental
J=0 states and the calculated states is less good for the general interaction than
for the state-dependent pairing interaction. 

 From Fig. 5 one can observe that the energy of the lowest EPC state is between the 1st and the
 second excited state (5). At $g=0$ this state corresponds to a pair moved from $g_{7/2}$ 
to $3s_{1/2}$. The occupation probabilities corresponding to the EPC state are
shown in Fig. 6(f). In the same figure are shown the occupation probabilities corresponding to
the 3rd SM state, which has the  closest resemblance to the structure of the EPC state. As
in the case of state-depending pairing interaction, the occupation probability for the state
$3_{1/2}$ is much larger for the EPC state than for the exact state.

 At high excitation energies the SM calculations with the general interaction  predict 
 a very large number of J=0 states. As illustrated in Fig. 5, this happens already for
 energies around 6 MeV. For this reason, it is difficult to search for the SM level which
 has the closest resemblance with a EPC state, as we have done for the pairing forces.

\section{Summary and Conclusions}

We have analysed the properties of excited states of zero seniority generated from the ground state
pair condensate.
One type of zero seniority states is obtained by breaking a pair from the pairing condensate and replacing it with an
"excited" pair. In addition, we have also considered a second type of  excited states which are  obtained by breaking
all the pairs from the ground state condensate and replacing them with identical excited pairs. The first and the second
type of zero seniority states are referred to as one-broken-pair and, respectively, excited pair condensate (EPC) states.
These states have been analysed  for the picket fence model and for the nucleus $^{108}$Sn. For this nucleus we have first 
performed a calculation with a state-dependent pairing interaction which is commonly used for tin isotopes. We found  that
the one-broken-pair states agree well with the known J=0 states in $^{108}$Sn. In the energy region of the one-broken-pair states there are many exact J=0 states, more than the experimentally known levels. Among the exact states one
can identify some states which are very similar in energy and structure to the one-broken-pair 
states. Based on these results, we concluded that  the experimentally known $J=0$
excited states in $^{108}$Sn are of one-broken-pair type.

For a state-independent pairing force we have first analysed the  EPC states which minimise the average
of the pairing Hamiltonian. We have identified two such EPC states, which appear at low energies, between
the first and the 3rd one-broken-pair states. In this energy region we did not find
exact states similar in structure to the two EPC states. However, at a much higher excitation energy,
of about 20 MeV, we have found an exact state which is similar in structure to an EPC state. This EPC state, 
denoted by EPC$^*$, is the reversed of the ground state condensate, in the sense that it is built on the
highest energy orbitals instead of the lowest one. It is shown that the EPC$^*$ state has the features 
of a pairing vibration state.

We have also analysed the seniority zero states in $^{108}$Sn for a general shell-model type 
interaction which was fitted to describe the low-lying states in tin isotopes. Compared with 
the previous case, the agreement between the  one-broken pair states and the experimentally knowns
J=0 states is better for the first level but less good for the other levels. In the region
of the high energy states there are many shell model states with J=0, which makes difficult
the comparison with the experimental data and also to the one-broken-pair states. For the general interaction
we found also a EPC state of relatively low excitation energy. Whether one could excite nuclei in such 
a particular low energy state, described by a pair condensate, is an interesting question which
deserves further studies.

\vskip 0.5cm
\noindent
{\bf Acknowledgements}
\vskip 0.2cm 

We thank  Daniel Negrea for participating to the initial phase of this study. 
N.S. is grateful for the hospitality of Institute of Modern Physics, Cantabria University, 
Spain, where this paper was mainly designed and written. A.P acknowledges a PhD fellowship from
the University of Bucharest. This work is supported by a grant of the Romanian Ministry of
Research and Innovation, CNCS - UEFISCDI, project number PCE 160/2021, within PNCDI III.

\section{Appendix}

In this Appendix we provide  the recurrence relations necessary to calculate the 
average of the pairing Hamiltonian in the states (4,5) as well as the norm of these
states.
We consider the case of spherically symmetric single-particle levels and we label
them by the standard quantum numbers $|n_il_ij_im_i> \equiv |i m_i> $ . We express the 
Hamiltonian in terms of spherically symmetric quantities, i.e.
\begin{equation}
	H= \sum_i^\Omega \epsilon_i N_{j_i} +
  \sum_{ij=1}^\Omega V_{ij} P^\dagger_{i} P_{j}
\end{equation}
where $V_{ij} = \sqrt{(2j_i+1)(2j_j+1)} <(ii)J=0|V|(jj)J=0>$. The operators $P^\dagger$
represent a pair of angular momentum J=0, i.e.
 
\begin{equation}
P_{i}^{\dagger}=\frac{1}{\sqrt{2j_i+1}}\sum_{m_i=-j_i}^{j_i}(-1)^{j_i-m_i}a_{m_i}^{\dagger}a_{-m_i}^{\dagger}
\end{equation}
In what follows the states (4,5) are denoted by 
\begin{equation}
|n;0 \rangle = (\Gamma^{\dagger})^n |0\rangle  
\end{equation}
\begin{equation}
|n;1 > = \tilde{\Gamma}^\dagger (\bar{\Gamma}^\dagger)^{N-1}| 0 >
\end{equation}
Since in the variational calculations it is involved only one state (5) at a time, 
in Eq. (16) we have removed the index $k$. With these notations the collective pair
operators have the expressions  
\begin{equation}
    \overline{\Gamma}^{\dagger}=\sum_i x_{i}P_{i}^{\dagger}
\end{equation}
\begin{equation}
	\tilde{\Gamma}^{\dagger}=\sum_i y_{i}P_{i}^{\dagger}
\end{equation}
Below we give the recurrence relations for the overlaps and for the operators which are 
involved in calculating the average of the pairing Hamiltonian. For all the recurrence relations
we provide also the initial quantities required to evolve them.

{\bf (a) Recurrence relations for the overlaps}
\begin{multline*}
 \langle n, 0 | n, 0 \rangle = n \sum_i x_{i}^2 \langle n-1,0 | n-1, 0 \rangle - 2n(n-1) \sum_i \frac{x^3_{i}}{2i+1} \langle n-2,0 | P_{i} | n-1, 0 \rangle
\end{multline*}
\begin{multline*}
 \langle n-1, 1 | n, 0 \rangle = (n-1) \sum_i x_{i}^2 \langle n-2,1 | n-1, 0 \rangle + 
\sum_i x_{i} y_{i} \langle n-1,0 | n-1, 0 \rangle \\
 - 2(n-1)(n-2) \sum_i \frac{x^3_{i}}{2i+1} \langle n-3,1 | P_{i} | n-1, 0 \rangle
 - 4(n-1) \sum_i \frac{x^2_{i} y_{i}}{2i+1} \langle n-2,0 | P_{i} | n-1, 0 \rangle
\end{multline*}
\begin{multline*}
 \langle n, 1 | n, 1 \rangle = n \sum_i x_{i} y_{i} \langle n-1,1 | n, 0 \rangle + \sum_i y^2_{i} \langle n,0 | n, 0 \rangle \\
 - 2n(n-1) \sum_i \frac{x^2_{i}y_{i}}{2i+1} \langle n-2,1 | P_{i} | n, 0 \rangle
 - 4n \sum_i \frac{x_{i} y^2_{i}}{2i+1} \langle n-1,0 | P_{i} | n, 0 \rangle
\end{multline*}
Initial quantities:
\begin{equation*}
 \langle 1,0|1,0 \rangle  = \sum_i x^2_{i}
\end{equation*}
\begin{equation*}
 \langle 1,0|0,1 \rangle  = \sum_i x_{i} y_{i}
\end{equation*}
\begin{equation*}
 \langle 0,1|1,0 \rangle  = \sum_i x_{i} y_{i}
\end{equation*}
\begin{equation*}
 \langle 0,1|0,1 \rangle  = \sum_i y^2_{i}
\end{equation*}
{\bf (b) Recurrence relations for the particle number operator $N_{i}$ }
\begin{equation*}
 \langle n, 0 | N_{i} | n, 0 \rangle = 2 n x_{i} \langle n-1,0 | P_{i} | n, 0 \rangle
\end{equation*}
\begin{equation*}
 \langle n-1, 1 | N_{i} | n, 0 \rangle = 2 n x_{i} \langle n-1,0 | P_{i} | n-1, 1 \rangle
\end{equation*}
\begin{equation*}
 \langle n, 1 | N_{i} | n, 1 \rangle = 2 n x_{i} \langle n-1,1 | P_{i} | n, 1 \rangle + 2 y_{i} \langle n,0 | P_{i} | n, 1 \rangle
\end{equation*}
Initial quantities:
\begin{equation*}
 \langle 1,0 | N_{i} | 1,0 \rangle = 2 x^2_{i}
\end{equation*}
\begin{equation*}
 \langle 1,0 | N_{i} | 0,1 \rangle = 2 x_{i} y_{i}
\end{equation*}
\begin{equation*}
 \langle 0,1 | N_{i} | 1,0 \rangle = 2 x_{i} y_{i}
\end{equation*}
\begin{equation*}
 \langle 0,1 | N_{i} | 0,1 \rangle = 2 y^2_{i}
\end{equation*}
{\bf (c) Recurrence relations for the pair operator $P_{i}$ }
\begin{multline*}
\langle n-1, 0 | P_{i} | n, 0 \rangle = n x_{i} \langle n-1, 0 | n-1,0 \rangle - \frac{2n(n-1)}{2i+1} x^2_{i} \langle n-2,0 | P_{i} | n-1, 0 \rangle   
\end{multline*}
\begin{multline*}
\langle n-2, 1 | P_{i} | n, 0 \rangle = n x_{i} \langle n-2, 1 | n-1,0 \rangle - \frac{2n(n-1)}{2i+1} x^2_{i} \langle n-2,0 | P_{i} | n-2, 1 \rangle
\end{multline*}
\begin{multline*}
\langle n-1, 1 | P_{i} | n, 1 \rangle = n x_{i} \langle n-1, 1 | n-1,1 \rangle + y_{i} \langle n-1, 1 | n,0 \rangle \\ - \frac{2n(n-1)}{2i+1} x^2_{i} \langle n-2,1 | P_{i} | n-1, 1 \rangle - \frac{4n}{2i+1} x_{i} y_{i} \langle n-1,0 | P_{i} | n-1, 1 \rangle    
\end{multline*}
Initial quantities: 
\begin{equation*}
 \langle - | P_{i} | 1,0 \rangle = x_{i}
\end{equation*}
\begin{equation*}
 \langle - | P_{i} | 0,1 \rangle = y_{i}
\end{equation*}
{\bf (d) Recurrence relations for the pairing interaction operator $P_i^\dagger P_j$ }
\begin{multline*}
	\langle n, 0 | P^{\dagger}_{i} P_{j} | n, 0 \rangle = n^2 x_{i} x_{j} \langle n-1,0 | n-1,0 \rangle - \frac{2n^2(n-1)}{2j+1} x_{i} x^2_{j} \langle n-2,0 | P_{j} | n-1,0 \rangle \\
 - \frac{2n^2(n-1)}{2i+1} x^2_{i} x_{j} \langle n-2,0 | P_{i} | n-1,0 \rangle  + \frac{2n(n-1)}{2i+1} \times \frac{2n(n-1)}{2j+1} x^2_{i} x^2_{j}[\langle n-2,0 | P^{\dagger}_{j} P_{i} | n-2,0 \rangle \\ + \delta_{ij} ( \langle n-2,0 | n-2,0 \rangle - \frac{2}{2i+1} \langle n-2,0 | N_{i} | n-2,0 \rangle )]
\end{multline*}
\begin{multline*}
	\langle n-1, 1 | P^{\dagger}_{i} P_{j} | n, 0 \rangle = n(n-1) x_{i} x_{j} \langle n-2,1 | n-1,0 \rangle - \frac{2n(n-1)^2}{2j+1} x_{i} x^2_{j} \langle n-2,0 | P_{j} | n-2,1 \rangle \\ + n x_{j} y_{i} \langle n-1,0 | n-1,0 \rangle - \frac{2n(n-1)}{2j+1} x^2_{j} y_{i} \langle n-2,0 | P_{j} | n-1,0 \rangle \\ - \frac{2n(n-1)(n-2)}{2i+1} x^2_{i} x_{j} \langle n-3,1 | P_{i} | n-1,0 \rangle
 + \frac{2(n-1)(n-2)}{2i+1} \times \frac{2n(n-1)}{2j+1} x^2_{i} x^2_{j}[\langle n-3,1 | P^{\dagger}_{j} P_{i} | n-2,0 \rangle \\ + \delta_{ij} ( \langle n-3,1 | n-2,0 \rangle - \frac{2}{2i+1} \langle n-3,1 | N_{i} | n-2,0 \rangle )] - \frac{4 n(n-1)}{2i+1} x_{i} x_{j} y_{i} \langle n-2,0 | P_{i} | n-1,0 \rangle \\ 
 + \frac{4(n-1)}{2i+1} \times \frac{2n(n-1)}{2j+1} x_{i} x^2_{j} y_{i} [\langle n-2,0 | P^{\dagger}_{j} P_{i} | n-2,0 \rangle + \delta_{ij} ( \langle n-2,0 | n-2,0 \rangle - \frac{2}{2i+1} \langle n-2,0 | N_{i} | n-2,0 \rangle)] 
\end{multline*}
\begin{multline*}
	\langle n, 1 | P^{\dagger}_{i} P_{j} | n, 1 \rangle =  n^2 x_{i} x_{j} \langle n-1,1 | n-1,1 \rangle + n x_{i} y_{j} \langle n-1,1 |n,0\rangle \\ - \frac{2n^2(n-1)}{2j+1} x_{i} x^2_{j} \langle n-2,1 | P_{j} | n-1,1 \rangle - \frac{4 n^2}{2j+1} x_{i} x_{j} y_{j} \langle n-1,0 | P_{j} | n-1,1 \rangle \\ + n x_{j} y_{i} \langle n,0 | n-1,1 \rangle + y_{i} y_{j} \langle n,0 |n,0\rangle
 - \frac{2n(n-1)}{2j+1} x^2_{j} y_{i} \langle n-2,1 | P_{j} | n,0 \rangle \\ - \frac{4n}{2j+1} x_{j} y_{i} y_{j} \langle n-1,0 | P_{j} | n,0 \rangle - \frac{2n^2(n-1)}{2i+1} x^2_{i} x_{j} \langle n-2,1 | P_{i} | n-1,1 \rangle \\
 - \frac{2n(n-1)}{2i+1} x^2_{i} y_{j} \langle n-2,1 | P_{i} |n,0\rangle 
 + \frac{4n^2(n-1)^2}{(2j+1)^2} x^2_{i} x^2_{j}[\langle n-2,1 | P^{\dagger}_{j} P_{i} | n-2,1 \rangle \\ + \delta_{ij} ( \langle n-2,1 | n-2,1 \rangle - \frac{2}{2i+1} \langle n-2,1 | N_{i} | n-2,1 \rangle )] \\
 + \frac{8n^2(n-1)}{(2j+1)^2} x^2_{i} x_{j} y_{j} [\langle n-2,1 | P^{\dagger}_{j} P_{i} | n-1,0 \rangle + \delta_{ij} ( \langle n-2,1 | n-1,0 \rangle \\ - \frac{2}{2i+1} \langle n-2,1 | N_{i} | n-1,0 \rangle)] - \frac{4 n^2}{2i+1} x_{i} x_{j} y_{i} \langle n-1,0 | P_{i} | n-1,1 \rangle \\ - \frac{4n}{2i+1} x_{i} y_{i} y_{j} \langle n-1,0 | P_{i} |n,0\rangle 
 + \frac{8n^2(n-1)}{(2j+1)^2} x_{i} x^2_{j} y_{i} [\langle n-1,0 | P^{\dagger}_{j} P_{i} | n-2,1 \rangle \\ + \delta_{ij} ( \langle n-1,0 | n-2,1 \rangle - \frac{2}{2i+1} \langle n-1,0 | N_{i} | n-2,1 \rangle)] \\
 + \frac{16 n^2}{(2i+1)^2} x_{i} x_{j} y_{i} y_{j} [\langle n-1,0 | P^{\dagger}_{j} P_{i} | n-1,0 \rangle + \delta_{ij} ( \langle n-1,0 | n-1,0 \rangle \\ - \frac{2}{2i+1} \langle n-1,0 | N_{i} | n-1,0 \rangle)]
\end{multline*}
Initial quantities:  
\begin{equation*}
 \langle 1,0 | P^{\dagger}_{i} P_{j} | 1,0 \rangle = x_{i} x_{j} 
\end{equation*}
\begin{equation*}
 \langle 1,0 | P^{\dagger}_{i} P_{j} | 0,1 \rangle = x_{i} y_{j} 
\end{equation*}
\begin{equation*}
 \langle 0,1 | P^{\dagger}_{i} P_{j} | 1,0 \rangle = x_{j} y_{i} 
\end{equation*}
\begin{equation*}
 \langle 0,1 | P^{\dagger}_{i} P_{j} | 0,1 \rangle = y_{i} y_{j} 
\end{equation*}

\end{document}